# Influence of small scale heterogeneity on CO2 trapping processes in deep saline aquifers

**Naum I. Gershenzon, Mohamad Reza Soltanian, Robert W. Ritzi Jr., David F. Dominic**

Department of Earth and Environmental Sciences, Wright State University, 3640 Col. Glenn Hwy., Dayton, OH 45435

Naum I. Gershenzon (naum.gershenzon@wright.edu)

**ABSTRACT**

The physical mechanism of CO2 trapping in porous media by capillary trapping (pore scale) incorporates a number of related processes, i.e. residual trapping, trapping due to hysteresis of the relative permeability, and trapping due to hysteresis of the capillary pressure. Additionally CO2 may be trapped in heterogeneous media due to difference in capillary pressure entry points for different materials (facies scale). The amount of CO2 trapped by these processes depends upon a complex system of non-linear and hysteretic relationships including how relative permeability and capillary pressure vary with brine and CO2 saturation, and upon the spatial variation in these relationships as caused by geologic heterogeneity.

Geological heterogeneities affect the dynamics of CO2 plumes in subsurface environments. Recent studies have led to new conceptual and quantitative models for sedimentary architecture in fluvial deposits over a range of scales that are relevant to the performance of some deep saline reservoirs. We investigated how the dynamics of a CO2 plume, during and after injection, is influenced by the hierarchical and multi-scale stratal architecture in such reservoirs. The results strongly suggest that representing small scales features (decimeter to meter), including their organization within a hierarchy of larger-scale features, is critical to understanding trapping processes.

# 1. INTRODUCTION

The idea of sequestering $CO_2$ in the Earth's crust has been discussed and evaluated for more than two decades (e.g. Steinberg, 1992; Hitchon et al., 1999; Bachu, 2000). The most likely sites for sequestration are deep saline aquifers and depleted hydrocarbon reservoirs (Holloway, 2001; Kovscek and Wang, 2005; Kovscek and Cakici, 2005, Bruant et al., 2002; Pruess and Garcı́a, 2002; Bachu, 2003; Deng et al., 2012; Dai et al., 2013). Sites must be evaluated for reservoir capacity and the risk of $CO_2$ leakage, which requires detailed modelling of $CO_2$ movement. Such modeling must account for coupled processes occurring over a wide range of scales. For example Middleton et al. (2012) defined and described processes relevant at these scales: sub-pore (Å -10 nm), pore (10 nm- 10 cm), $CO_2$ reservoir (10 cm – 100 m), site (100 m – 10 km, the deep groundwater system), and region (10 km – 1 Mm, the sedimentary basin). At any site dependent on structural (or hydrodynamic) trapping, caprock may be compromised by improperly abandoned wells, stratigraphic discontinuities, or faults. Therefore, other trapping mechanisms must be considered, including, dissolution, mineralization, and capillary trapping. We focus on capillary trapping processes that operate over a range of scales of heterogeneity intermediate between the pore scale (c.f. Kuo and Benson, 2013) and the site scale (c.f. Zhou et al., 2009), scales that have not yet received full consideration in the literature.

Typically, $CO_2$ is injected in a supercritical state with liquid-like properties. After injection the $CO_2$ plume migrates towards the top of the reservoir, driven by buoyancy forces. $CO_2$ becomes virtually immobile due either to capillary trapping in reservoir rock pores (time scale: 10 to 100 years), dissolution of $CO_2$ into brine (100 to 1000 years), or mineral trapping through reaction with rock minerals (thousands to millions of years) (Xu et al., 2003; IPCC, 2007; and Hesse et al., 2007). We modelled both capillary trapping and dissolution processes, but our main interest is in capillary trapping, especially the effects of reservoir heterogeneity.

Geological heterogeneities affect the dynamics of $CO_2$ plumes in subsurface environments. Their role in capillary trapping and dissolution of $CO_2$ has been investigated extensively (Hovorka et al., 2004; Doughty and Pruess, 2004; Kumar et al., 2005; Bryant et al., 2006; Forster et al., 2006; Ambrose et al., 2008; Ide et al., 2007; Flett et al., 2007; Tsang et al., 2008; Han et al., 2010; Sakaki et al., 2013). Usually the effects of heterogeneity are considered within two-dimensional geostatistical models using various correlation lengths. While such an approach is capable of capturing qualitatively some typical features of the process, full understanding requires modeling three-dimensional flow within reservoirs with realistic representations of sedimentary architecture and the associated relative permeability and capillary pressure distributions, across a range of scales.

Recent studies have led to new conceptual and quantitative models for sedimentary architecture in fluvial deposits over a range of scales that are relevant to fluid flow in some petroleum and deep saline reservoirs (e.g. Lunt et al., 2004; Bridge, 2006; Lunt and Bridge 2007). From these studies emerged a generalized, three-dimensional, quantitative model for the multiscale and hierarchical stratal architecture found in fluvial deposits. Importantly, the lengths of stratal units at all hierarchical levels scale together with the width of the formative channels (Bridge, 2006), making it possible to adapt the general model to specific deposits. The software package (GEOSIM) uses a geometric-based approach to create three-dimensional geocellular models representing this multiscale and hierarchical fluvial architecture (Ramanathan et al., 2010; Guin et al. 2010; Ritzi, 2013).

Here we focus on fluvial deposits dominated by sandy gravel. Lunt et al. (2004) studied the gravelly channel belt of the Sagavanirktok River Alaska, an analog for a number of important reservoirs composed of fluvial deposits (e.g. Tye et al., 2003, Zhou et al., 2009). At the smallest scale are sets of cross-strata (decimeters thick and meters long), which occur within unit bar deposits (tens of decimeters thick and tens of meters long). Unit bars and cross-bar channel fills occur within compound bar deposits (meters thick and hundreds of meters long). Compound bar deposits and the channel fills that bound them occur within channel belts (tens of meters thick and kilometers long). Importantly, open-framework gravel (OFG) cross strata are known to create preferential flow pathways that confound attempts at gas injection (e.g. Tye et al., 2003). The OFG are found to make up 25 to 30 percent of the volume of a deposit.

The complexity of the heterogeneity and highly non-linear nature of the problem make it challenging to achieve numerically convergent solutions. Consequently, the simulations presented here are limited to an examination of CO2 within a relatively small reservoir compartment of 100 m x 100 m x 5 m (illustrated in Gershenzon et al., 2014). Our longer-term goal is to run larger simulations that include more of the geocellular model (e.g. Ramanathan et al., 2010, Figure 10), and thus include some of the larger-scale sedimentary architecture not yet represented here. However, the preliminary work presented here shows the fundamental importance of properly representing the within-reservoir heterogeneity, and therefore justifies this line of research.

In the following sections we review background research (Section 2), give short overview of fluvial reservoir model (Section 3), describe the simulations framework of CO2 sequestration in saline aquifer (Section 4) and finally present and discuss the results of simulation (Section 4).

## 2. BACKGROUND

Here we will focus our attention mainly on processes at the reservoir scale. The main short-term (~10 years) physical mechanism of CO2 trapping in porous media is capillary trapping (this term incorporates three related mechanisms, i.e. residual trapping, trapping due to hysteresis of the relative permeability, and trapping due to hysteresis of the capillary pressure) (Kumar et al., 2005; Mo et al., 2005; Spiteri et al., 2008; Juanes et al., 2006; Altundas et al., 2011). The basics of this mechanism are as follow. After injection into deep saline aquifers, the low viscosity CO2 tends to migrate to the top of the geologic structure due to a density difference between the CO2 and the brine. During the injection period CO2 displaces brine in a drainage process. After the injection is finished, the buoyant CO2 migrates upward and water displaces CO2 at the plume "tail" in an imbibition-like process. The latter causes CO2 stream disconnection into immobile blobs and ganglia (Hunt et al., 1988). The amount of CO2 trapped by this mechanism is a complicated nonlinear function of the spatial distribution of permeability, permeability anisotropy, capillary pressure, relative permeability of brine and CO2, permeability hysteresis and residual gas saturation, as well as the rate, total amount and placement of injected CO2. Kumar et al. (2005), analyzing residual and aqueous CO2 trapping, showed that the former is a dominant short-term trapping mechanism. Aquifer dip and permeability anisotropy have an effect on gas migration and hence affects CO2 dissolution in brine. Juanes et al. (2006) found that relative permeability hysteresis becomes a major factor in CO2 trapping. High injection rates and injection of water slugs alternating with CO2 injection increases the effectiveness of the sequestration project. They also observed that coarse simulation models overestimate capillary trapping of CO2, suggesting high-resolution models for accurate assessment of the different storage mechanisms. Ide et al. (2007) established that the amount of CO2 immobilized by residual trapping is a strong function of the gravity number (the ratio between viscous and gravitational forces (Zhou et al. 1994). Altundas et al. (2011) showed that not only hysteresis of permeability but also hysteresis of capillary pressure affects the dynamics of CO2 trapping. The latter mechanism works not only on the tail of the CO2 plume but at the front as well, causing its retardation.

The solubility trapping is also an important and safe CO2 storage mechanism (Pruess and García, 2002; Ennis-King and Paterson, 2005). CO2 dissolution is a slow diffusion process. The rate of CO2 dissolution in brine depends on the contact area between these two phases. Since the CO2-brine solution has a density greater than brine alone, convective motion is initiated under appropriate conditions (Lindeberg and Wessel-Berg, 1997). If convection does develop, the contact area considerably increases, and this mechanism may become a dominant form of CO2 storage in aquifers over periods of tens to hundreds of years.

Numerous studies have considered the influence of different types of heterogeneity on reservoir performance of CO2 sequestration in saline formations. Hovorka et al. (2004), modeling a CO2 injection test in the Frio Formation of the upper Texas Gulf Coast, concluded that stratigraphic heterogeneity may result in significant channeling of the flow of a CO2 plume. Doughty and Pruess (2004) created three-dimensional models representing the fluvial/deltaic Frio Formations in the upper Texas gulf coast. Their simulations also demonstrate the strong influence of geologic heterogeneity on buoyancy-driven CO2 migration. Obi and Blunt (2006), considering the lateral movement (advection) of the CO2 plume under a horizontal pressure gradient, found that permeability heterogeneity causes the CO2 to migrate much further than predicted from a homogeneous model. This allows trapping mechanisms to be more efficient than in the homogeneous case, leading to significant impedance of the movement of $CO_2$. However the sequestration efficiency of reservoir is considerably less than in the homogeneous case. Ide et al. (2007) studied $CO_2$ plume dynamics in two-dimensional layered reservoirs. The permeability maps for the various layers were created using a geo-statistical model. The influences of different correlation coefficients on the permeability distribution were analyzed. It was shown that the difference is small in the gas distribution between the homogeneous case and the cases of heterogeneity with short and long permeability correlation. Bryant et al. (2006) investigated the instability of the $CO_2$-brine interface in heterogeneous two-dimensional reservoirs. They observed the formation of narrow closely-spaced $CO_2$ channels. The placement of these channels is correlated with permeability channels, suggesting that the preferential–flow paths of buoyancy-driven $CO_2$ displacement are due to heterogeneity of the aquifer rather than fingering due to viscous instability. The rising $CO_2$ front is smoothed if capillary pressure is included in the simulation, since capillary and buoyancy forces are of similar magnitude. Bryant et al. (2006) and Ide et al. (2007) showed that a layer of smaller permeability lying above a layer of larger permeability can act as a seal due to capillary pressure barrier. Saadatpoor et al. (2009) further exposed this additional trapping mechanism, which they termed a "local trapping mechanism." Supposing a functional dependence between permeability and capillary pressure, they simulated the dynamics of the buoyant $CO_2$-brine front in a heterogeneous reservoir. Comparison of this simulation with an analogous simulation but with a homogeneous (averaged) capillary pressure curve revealed a dramatic difference in results. In the former case $CO_2$ rises through the high permeability channels, which are surrounded by the capillary barriers of the low permeability material. In some regions capillary barriers prevent upward movement of $CO_2$, allowing only lateral migration, which effectively traps the $CO_2$ plume. The same idea has been exploited by Zhou et al. (2009). Considering a regional-scale (typical lateral size up to a few km) flow and transport processes in a layered reservoir, they investigated the "secondary-seal effect", i.e. the effect of $CO_2$ accumulation between layers with different permeability and capillary pressure entry points, causing retardation of upward $CO_2$ migration. Flett et al. (2007), using standard geo-statistical techniques, generated various three-dimensional reservoir realizations with different sand-to-shale proportions. They found that the increase of shale content reduces the rate of residual $CO_2$ trapping. Han et al. (2010) systematically investigated the influence of heterogeneity in the permeability distribution (generated by geo-statistical techniques in two-dimensional reservoirs) on $CO_2$ trapping mechanisms and buoyancy-driven $CO_2$ migration. They found that 1) the amount of residually trapped $CO_2$ is enhanced if $CO_2$ plume migrates farther in the vertical or lateral direction; 2) the presence of low permeability inclusions (secondary caprock) enhances residual $CO_2$ trapping; 3) increasing the degree of permeability heterogeneity decreases the amount of residual $CO_2$ trapping.

Simulations at the reservoir scale critically depend on relative permeability and capillary pressure curves for drainage and imbibition (e.g. Doughty and Pruess, 2004; Kumar et al., 2005; Mo and Akervoll, 2005; Juanes et al., 2006; Burton et al 2009; Altundas et al., 2011). These parameters are difficult to measure and are quite uncertain for most potential reservoirs. Various approaches have been used to determine interconnection between petrophysical parameters and to specify the Brooks-Corey relations for capillary pressure and relative permeability curves (Holtz, 2002; Altundas et al., 2011; Delshad et al., 2013). To define irreducible water saturation and residual $CO_2$ saturation, Holtz (2002) uses empirical relations between these parameters and porosity, and the latter is defined from the empirical relation between permeability and porosity. Altundas et al. (2011) modified the Brooks-Corey relation for imbibition by redefining the normalization of relative saturation and introducing a coefficient to characterize the pore-body radius to pore-throat radius ratio. Delshad et al. (2013) have proposed a method that uses the relations between petrophysical parameters and interfacial tension to define the permeability and capillary pressure. Krevor et al. (2012) measured capillary pressure and relative permeability for four different sandstone rocks including Mt. Simon and Berea formations (the potential sites for $CO_2$ injection). They provide the best-fit Brooks-Corey parameters for both capillary pressure and relative permeability curves for $CO_2$.

Different storage strategies have been proposed to reduce the amount of mobile $CO_2$ and to increase the capacity of reservoir. The "inject low and let rise" strategy is the obvious one (the larger distance between an injection point and a caprock the more $CO_2$ will be trapped by capillary trapping) (Kumar et al, 2005). Juanes et al. (2006) showed that high injection rates and a shorter injection period as well as injection of a water slug alternating with CO2 injection result in more effective sequestration. The simulation results by Ide et al. (2007) suggest that the amount of trapping can be maximized by injecting CO2 at the lowest feasible value of the gravity number, which can be controlled by adjusting the injection rate. Leonenko and Keith (2008) proposed injecting brine after CO2 to dissolve all of injected CO2 in a shorter time. The downside of this strategy is a severe limitation of storage capacity of an aquifer. Qi et al. (2009) demonstrated that simultaneously injected CO2 and formation brine followed by chase brine mitigates the mobility contrast between injected and displaced fluids, leading to higher storage efficiencies than injecting CO2 alone. Shamshiri and Jafarpour (2012), based on simulations in heterogeneous reservoirs and counting CO2 capillary trapping and solution in brine, proposed optimization algorithms to improve the CO2 storage efficiency.

## 3. OVERVIEW OF THE GEOLOGIC MODEL

The complexity and uncertainty of stratal architecture of saline aquifers and hydrocarbon reservoirs make it difficult to predict the actual behavior of a CO2 plume and to evaluate the reservoir capacity and the risk of CO2 leakage. Many sites suitable for CO2 sequestration have fluvial-type architecture, i.e. structure with multi-scale heterogeneities ranging from the centimeter to the reservoir scale.

The sedimentary architecture of reservoirs created by fluvial deposition has recently been described in three-dimensional models (Bridge, 2006). As shown in Figure 1, channel-belt deposits are characterized by a large volume fraction of convex-up, bar deposits formed within channels. Only when channels are abandoned and filled under lower flows are "channel-shaped" units formed. These concave-up channel fills are low-permeability baffles within the domain. In gravelly channel-belt deposits, preferential flow pathways arise from the interconnection of open-framework gravels within lobate unit bar deposits (Lunt and Bridge, 2007). These are the "thief zones" known by their negative effect on oil recovery (McGuire et al., 1999; Tye et al., 2003).

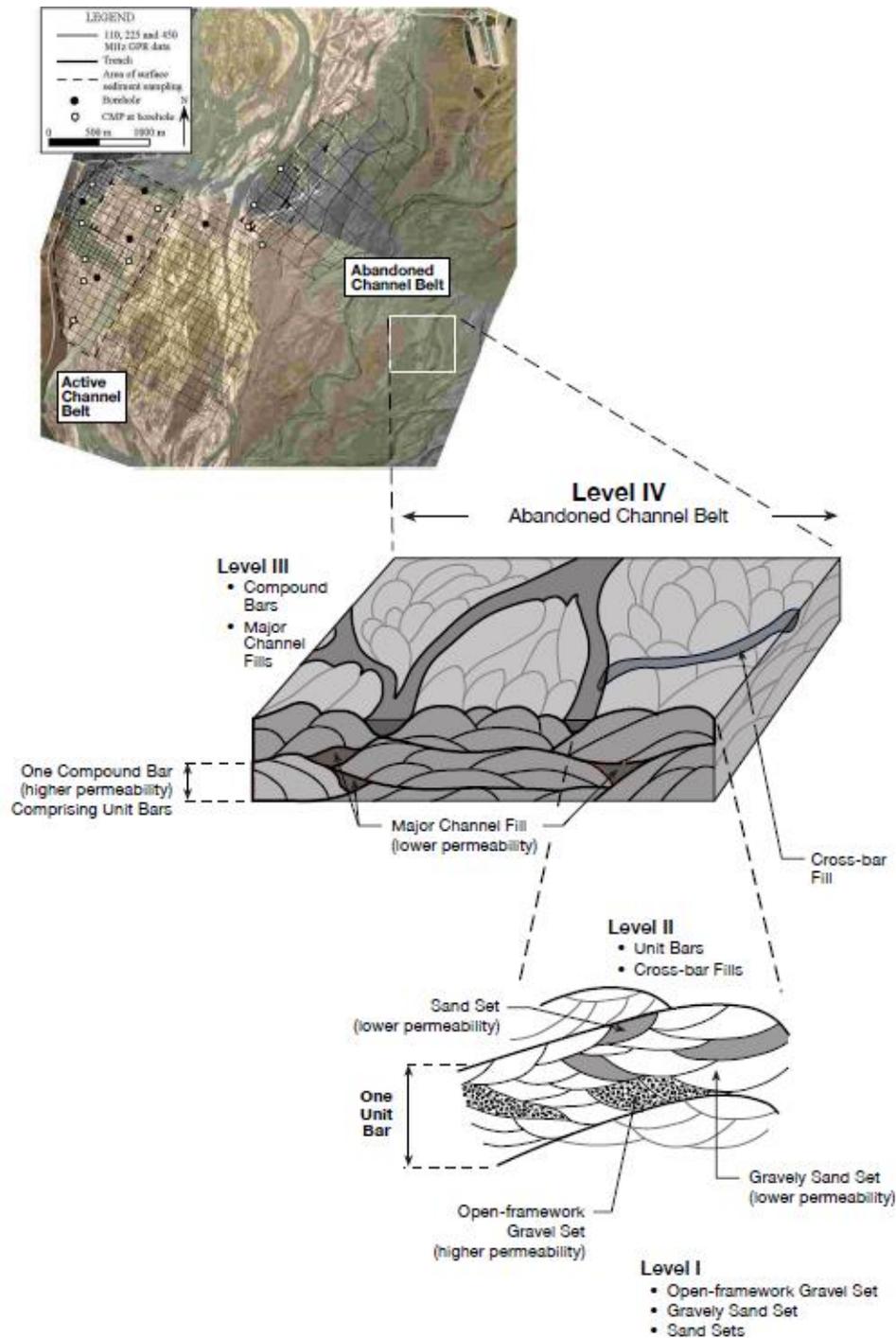

**Figure 1**. (Top) Study areas in the active channel belt and in the preserved channel belt deposits of the Sagavanirktok River (Lunt et al., 2004). (Middle and Bottom) Conceptual model for the hierarchical sedimentary architecture found in channel belt deposits (see also Table 1). The compound bar deposits at level III result from the processes of unit-bar accretion and channel migration. Within unit bar deposits (level II), sets of open-framework gravel (level I) have highest permeability. As channels are abandoned, they are filled with lower-permeability sediment. Major channel fills (level III) and smaller cross-bar channel fills (level II) are lower-permeability baffles within the deposit. From Ramanathan et al. (2010).

Lunt et al. (2004) studied the gravelly channel belt of the Sagavanirktok River (Alaska, Figure 1) and quantified the proportions and lengths for unit types across relevant scales (Table 1). At the smallest scale, sets of cross-stratified ("cross-sets" herein) sand, sandy gravel, and open-framework gravel (decimeters thick and meters long) occur within unit bar deposits (tens of decimeters thick and tens of meters long). Unit bars and cross-bar channel fills occur within compound bar deposits (meters thick and hundreds of meters long). Compound bar deposits and the channel fills that bound them occur within channel belts (tens of meters thick and kilometers long). Importantly, the open-framework gravel cross-sets were found to make up 25 to 30 percent of the volume of the deposit.

Table 1. Hierarchy of Unit Types

| IV | channel-belt deposit | | | | |
|---|---|---|---|---|---|
| III | compound bar deposits[1] | | | | major channel fills |
| II | unit bar deposits | | | cross-bar channel fills | concave-up sand |
| I | open-framework gravel set[2] | gravelly sand set | sand set | concave-up sand | concave-up sand |

[1.] typical dimensions (largest unit type): *750 x 500 x 2 $m^3$*
[2.] typical dimensions (smallest type): decimeters to meters long and wide, centimeters to decimeters thick

The sedimentary architecture quantified by Lunt et al. (2004) was incorporated by Ramanathan et al. (2010) into a high-resolution, three-dimensional, digital model using geometric-based simulation methods. Guin et al. (2010) confirmed that the model contains the hierarchy of sedimentary unit types and honors the proportions, geometries, and spatial distribution of the unit types quantified at each level by Lunt et al. (2004). Figure 2 shows a cross section through an extracted piece of a simulated compound bar deposit sampled with fine resolution (larger scale simulations showing larger-scale architecture are given in Ramanathan et al. (2010, Figure 10).

In this model, cross-sets of high-permeability open-framework gravel (OFG) were simulated discretely. OFG cells are considered to be connected when cell faces are adjacent. Importantly, clusters of continuously connected OFG cells create preferential flow pathways. When OFG cross-sets comprise at least 20% of the volume of the deposit, clusters span opposing pairs of domain boundaries (Guin et al., 2010). Such spanning, preferential-flow pathways exist as "thief zones" and control miscible gas injection within the Ivishak Formation (Tye et al., 2003).

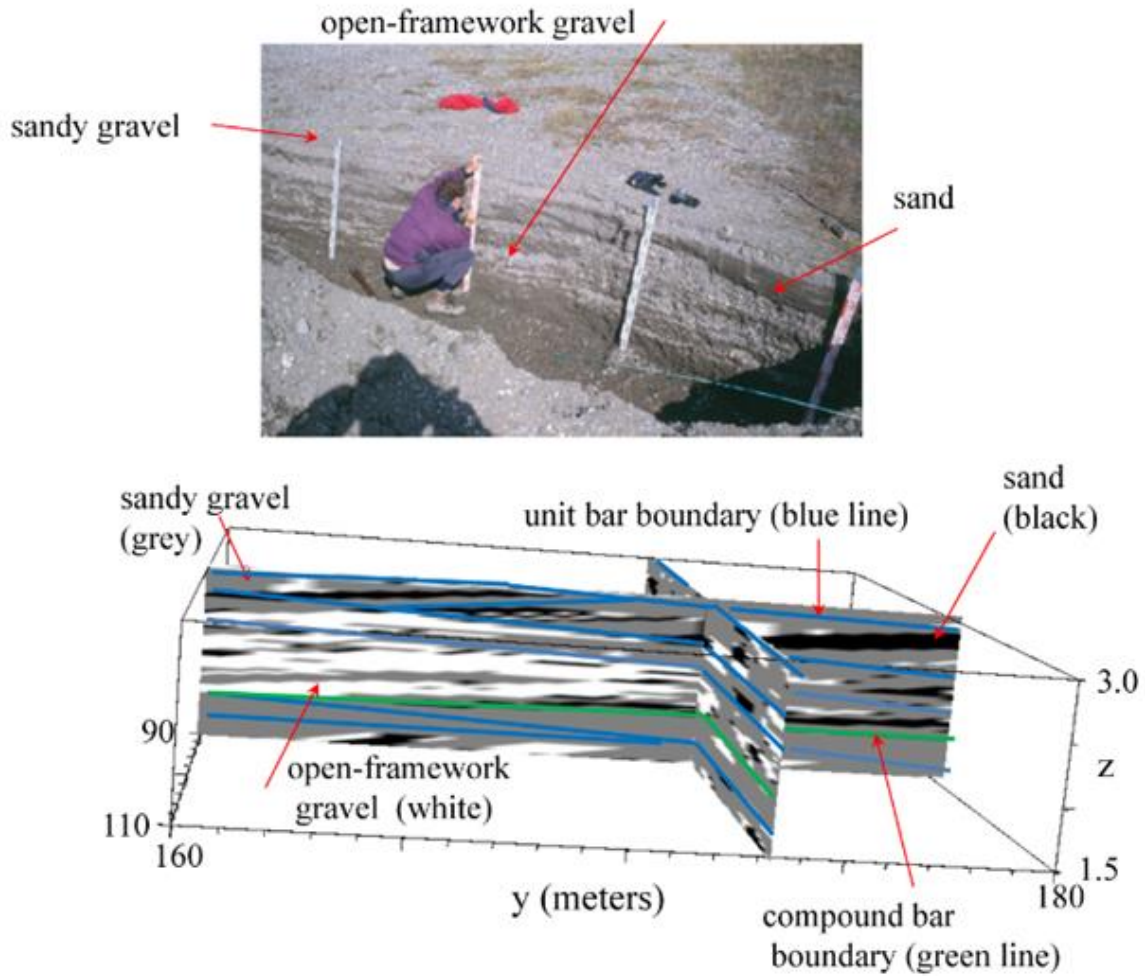

**Figure 2.** (Top) Exposure of level I unit types in a trench at the Sagavanirktok River field site (from Lunt, 2002). (Bottom) Rendering of orthogonal sections through an extracted piece of the stratal model produced for realization 1 with the GEOSIM code. The extracted piece was chosen so that open-framework gravel/conglomerate (28% of overall model volume) was clearly visible. Paleoflow direction is to the left.

The number, size, and orientation of OFG clusters in the model change with proportion. At any given proportion, the number, size, and orientation of clusters changes across the different hierarchical levels (scales) of the stratal architecture. Connected OFG cells within individual cross-sets form paths that vertically span single unit bar deposits. Connections across unit bar boundaries enhance lateral branching and the many clusters within unit bars connect into a smaller number of larger clusters at the scale of multiple unit bars. At the scale of a whole compound bar deposit, these clusters are typically connected into one or two large, spanning clusters. The spanning clusters occur at proportions of open-framework gravel cells below the theoretical threshold (31%) for random infinite media predicted in the mathematical theory of percolation. Guin and Ritzi (2008) showed that this is caused by geological structure within a finite domain. The percolation theory shows why two-dimensional models under-represent true three-dimensional connectivity, and thus why simulations must be three-dimensional (Huang et al., 2012).

The saturated permeability in sandy-gravel deposits varies non-linearly as a function of the volume of sand mixed with gravel (see Figure 6 in Ramanathan et al., 2010). Sandy-gravel strata have permeabilities similar to the sand they contain, which are of the order of $10^0$ to $10^1$ Darcies. Thus, sand and sandy-gravel cross-sets within unit bars have permeabilities similar to channel-fill sands. Open-framework gravels have permeabilities of the order of $10^3$ to $10^4$ Darcies. In either type of strata, the coefficient of variation in permeability is of the order of unity. In the lithified stratatypes within the Ivishak Formation (sandstones, pebbly sandstones and open-framework conglomerates), the saturated permeabilities scale down accordingly (Tye et al., 2003).

## 4. METHODOLOGY

Using a code by Ramanathan et al. (2010) We generated a realization of the channel-belt architecture including two materials, i.e. sand (76%) and OFG (24%) with geometric mean permeability 58 mD and 3823 mD, respectively. We simulated injection of $CO_2$ in a reservoir with size 100 m x 100 m x 5 m (250 thousand cells of size 2 m x 2 m x 0.05 m). This represents the heterogeneity created by an assemblage of unit bars within a compound bar. This simulation includes a cluster of OFG cells which tortuously spans the domain boundaries in all directions and contains 55% of all OFG cells. $CO_2$ was injected at a rate of 3.6 (standard) $m^3$/day during 10 days into the bottom of a vertical well at a depth 2360 m. In one case the well was placed to penetrate a spanning OFG cluster and in another case placed so it does not. The boundary of the $CO_2$ reservoir was not permeable. For comparison we also simulated injection into two homogeneous reservoirs: (1) homogeneous, isotropic, and with permeability equal to the geometric mean of the heterogeneous reservoir; and (2) homogeneous but anisotropic with permeability five times smaller in the z direction than in horizontal directions.

When simulating CO2 injection in a reservoir with realistic heterogeneity, the choices for relative permeability and capillary pressure tables are very important. The tables define the relations between water relative permeability, CO2 relative permeability, and CO2 capillary pressure as a function of water saturation. Two different sets of tables were utilized for the sand and OFG unit types. Thus, the total number of property tables was 12 including six for drainage and six for imbibition.

To generate tables we used the following methodology. First we find the irreducible water saturation and the maximum CO2 saturation, adapting the approach described by Holtz (2002). Using an empirical relation between permeability ($k$) and porosity $\emptyset$, we obtain porosity for sand and OFG (see Table 2).

$$k = 7 \cdot 10^7 \emptyset^{9.61}. \tag{1}$$

Then from the following empirical relation (Holtz, 2002) we find the averaged irreducible water saturation ($S_{wi}$) for both materials (Table 2).

$$S_{wi} = 5.159(\log(k)/\emptyset)^{-1.559} \tag{2}$$

Residual CO2 saturation ($S_{gr}$) can be found from the following empirical relation (Holtz, 2002):

$$S_{gr} = -0.969\emptyset + 0.5473 . \tag{3}$$

The next step is calculation of capillary pressure for drainage ($P_{cd}$) using the Brooks-Corey relation (Brook and Corry, 1964):

$$P_c = P_e \left(\frac{S-S_{wi}}{1-S_{wi}}\right)^{-1/\lambda}, \tag{4}$$

where $S$ is water saturation, $P_e$ is the minimum pressure required for the entry of CO2 into the pore of the rock and $\lambda$ is a fitting parameter known as pore size distribution index. For rocks from the Mt. Simon Formation $P_e = 4.6 \cdot 10^3 Pa$ and $\lambda = 0.55$ (Krevor et al., 2012). We use this value for the sand. To find the entry pressure point for OFG we scale the sand value based on the Leverett *J*-function as Saadatpoor et al. (2009) proposed:

$$P_e^{OFG} = P_e \left(\frac{k_{sand}\emptyset_{OFG}}{k_{OFG}\emptyset_{sand}}\right)^{0.5}.$$

For imbibition we use a curve analogous to (4) but, in contrast to drainage, the imbibition curve cross the saturation axis at value $S = 1 - S_{gr}$.

For the relative permeability curves the Brooks-Corey relations in the form proposed by Dullien (1992) are used:

$$k_{r,w} = (S_w^*)^{N_w} \tag{5}$$

$$k_{r,CO2} = k_{r,CO2}(S_{wi})(1 - S_w^*)^2[1 - (S_w^*)^{N_{CO2}}], \tag{6}$$

where $S_w^* = \frac{S-S_{wi}}{1-S_{wi}}$ for (5) and $S_w^* = \frac{S-S_{wi}}{1-S_{gr}-S_{wi}}$ for (6). The variables $N_w$ and $N_{CO2}$ are used as fitting parameters known as the Corey exponents for water and CO2, respectively. The following values are used: $N_{CO2} = 4$ (Krevor et al., 2012), $N_w = 5$ and $N_w = 3$ for drainage and imbibition. The values of other parameters are in the Table 2.

Table 2

|  | $\emptyset$ | $S_{wi}$ | $S_{gr}$ | $k_{r,CO2}(S_{wi})$ | $P_e$ (Pa) |
|---|---|---|---|---|---|
| sand | 0.23 | 0.22 | 0.32 | 0.65 | $4.6 \cdot 10^3$ |
| OFG | 0.36 | 0.1426 | 0.2 | 0.95 | $0.72 \cdot 10^3$ |

The results are summarized in Figure 3.

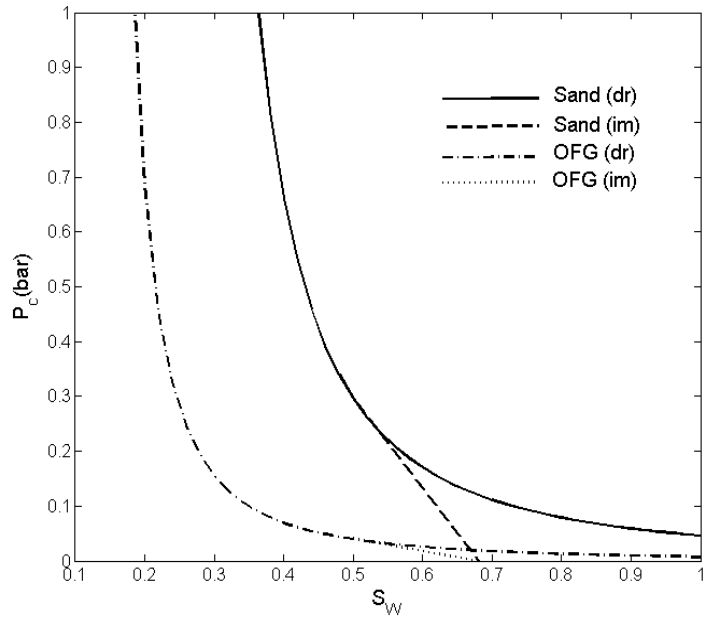

Figure 3a. Capillary pressure for sand and OFG for drainage and imbibition.

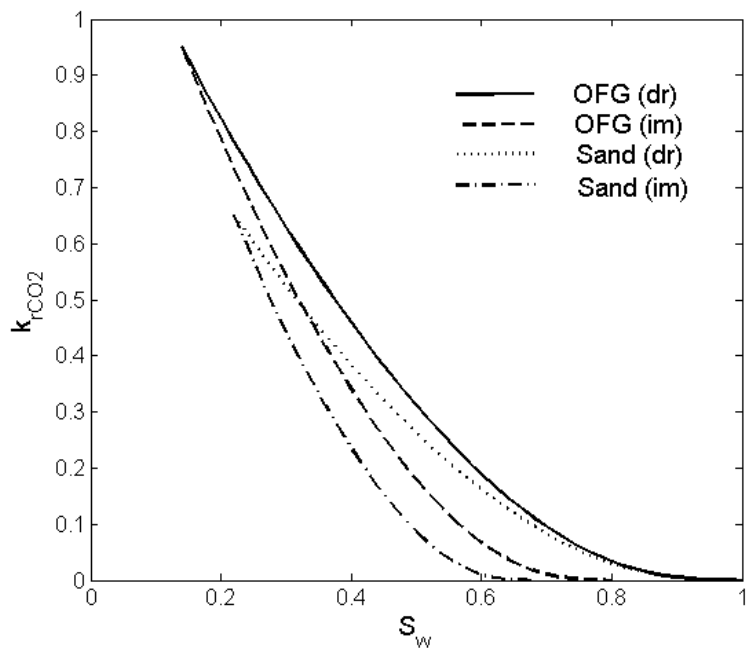

Figure 3b. Relative permeability of CO2 in sand and OFG for drainage and imbibition.

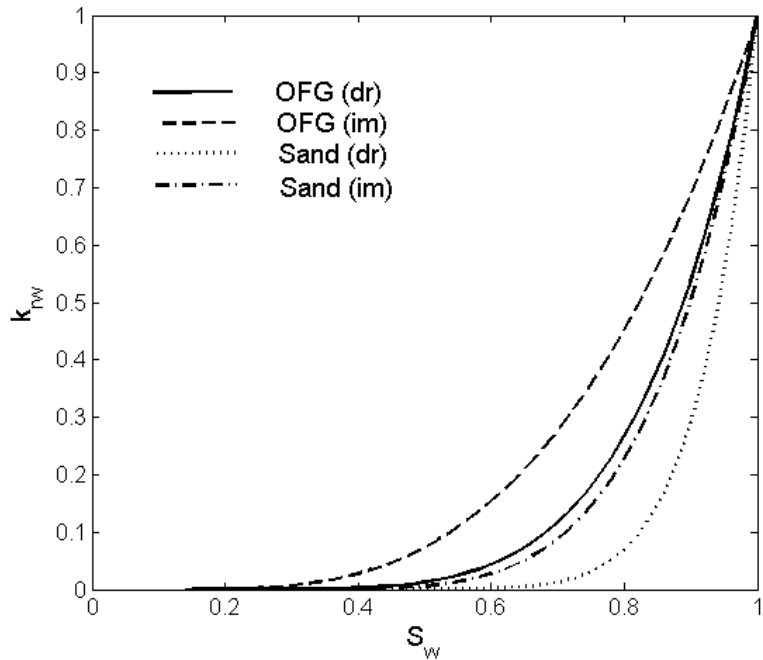

Figure 3c. Relative permeability of water in sand and OFG for drainage and imbibition.

## 5. RESULTS AND DISCUSSION

Figures 4 and 5 shows vertical cross sections with CO2 saturation for the heterogeneous and homogeneous anisotropic reservoirs after 10 days and after 1010 days, respectively. After injection CO2 propagates in the horizontal and vertical directions. The force of buoyancy moves the CO2 plume up.

The ratio of the viscous to gravity forces defines the geometry of the plume in the homogeneous case. In the homogeneous anisotropic reservoir the CO2 plume reaches the top of the reservoir in 10 days and then the top of the plume slowly spreads laterally. Essentially the same behaviour is exhibited by a plume in the homogeneous isotropic reservoir (not shown), but in the latter the plume reaches the top of the reservoir 5 times faster (in 2.5 days). The shape of the plume, the rate of capillary trapping of CO2, and the CO2 solution in brine are similar in these two cases.

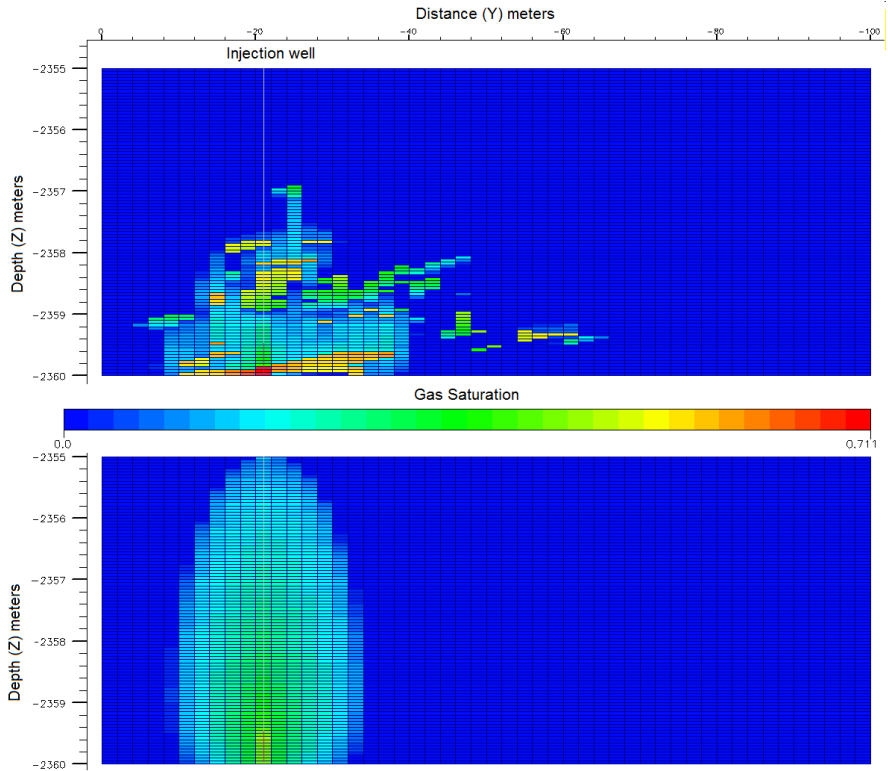

Figure 4. Vertical cross section of heterogeneous (top) and homogeneous anisotropic (bottom) reservoirs showing CO2 saturation after 10 days from the beginning of CO2 injection.

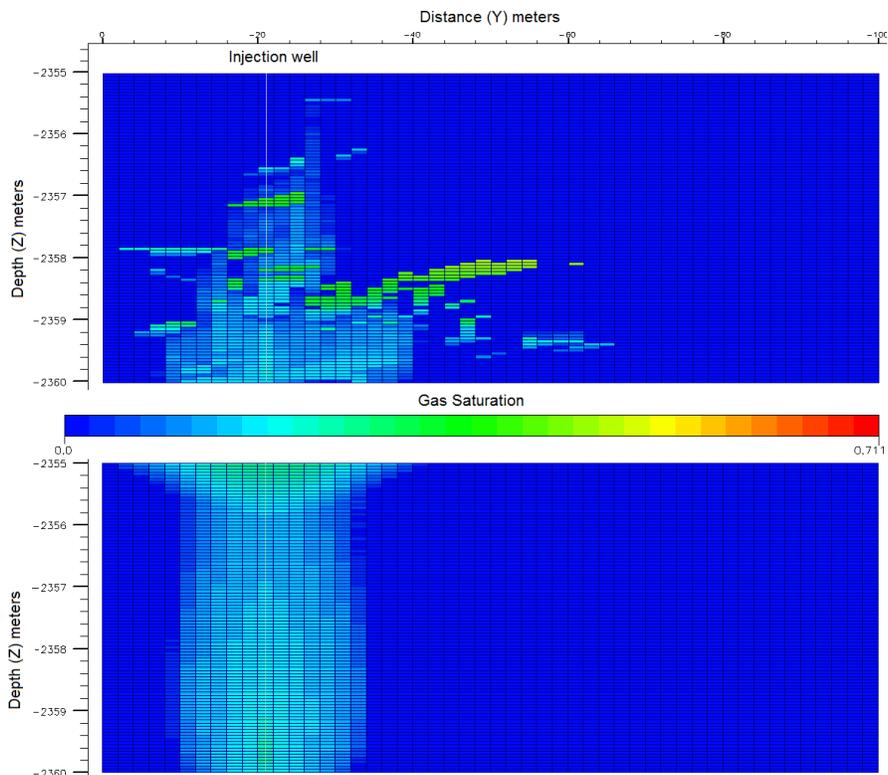

Figure 5. Vertical cross section of heterogeneous (top) and homogeneous (bottom) reservoirs showing CO2 saturation after 270 days from the beginning of CO2 injection.

In contrast, in the heterogeneous case the plume is two times wider and never reaches the top of the reservoir (see Figure 5, top panels). This is caused by the difference in capillary entry pressure between sand and OFG and the presence of multiple boundaries between sand and OFG. After injection a buoyancy force moves the $CO_2$ plume up. In the heterogeneous case it is more complicated. $CO_2$ propagates faster in OFG clusters since permeability of OFG material is much higher than in sand. However, in clusters that do not span the domain, $CO_2$ cannot escape from the cluster and remains trapped unless the buoyancy force is large enough to overcome the capillary entry pressure of sand. In spanning clusters, $CO_2$ propagates mainly in the horizontal direction since OFG clusters extend further in this direction. $CO_2$ can exit from a spanning OFG cluster through its upper boundary if buoyancy force is large enough. It will propagate upward through sand to the next OFG cluster, into which it will be pushed by both buoyancy force and capillary pressure. This process continues up to the point when buoyancy force becomes comparable with capillary pressure and the plume becomes immobile or trapped, or the plume reaches the top of the reservoir.

Due to the process described above the contact area between brine and the $CO_2$ plume is larger than in the homogeneous case. As a result the dissolution rate is larger in the heterogeneous case. The "regular" capillary trapping rate is also largely affected. Overall the results indicate that the presence of small and large OFG clusters essentially controls behavior of $CO_2$ plumes on reservoir scale.

Importantly, in the heterogeneous case considered here the plume never reaches the caprock. Total amount of inserted $CO_2$ is effectively trapped. The trapping is mostly in sand (in blobs and ganglia) and by the secondary sealing effect in OFG material (on the boundary between sand and OFG).

Injected $CO_2$ can be apparently divided into four parts, i.e. 1) mobile gas, 2) immobile gas trapped in the form of blobs and ganglia by hysteresis, 3) gas trapped at the boundary between sand and OFG due to different entry point pressures, and 4) dissolved gas. Note that gas trapped at the boundary between sand and OFG could be mobilized by increased gas injection, pushing it through the boundary.

Figure 6 depicts the total amount of the mobile, capillary trapped, and solute CO2 in sand and OFG. While OFG material comprises only 24% of the reservoir, most of the mobile CO2 is in OFG cells. (Figure 6a). The amount of capillary trapped CO2 in OFG is about twice that in sand (Figure 6b). The amount of solute CO2 is two to three times larger in sand than in OFG (Figure 6c).

To further illustrate the roles of 1) capillary pressure, 2) hysteresis of relative permeability of water and CO2 and 3) heterogeneity of permeability on plume geometry and dynamics as well as on CO2 trapping and dissolution we compare the results of the five cases described in Table 3.

The results from each case are given in Figure 7. Case 1 includes heterogeneity and all 12 property tables; it is the standard for comparison. This figure shows that plume geometry is very different from case to case. The plume quickly reaches the top of the reservoir for the homogeneous reservoir (case 2), in heterogeneous reservoir with capillary pressure "off" (case 5) and if capillary pressure is "on" but the properties tables are the same for both materials (case 3). In contrast, the spreading of the plume in the vertical direction is much slower in cases 1 and 4. The results illustrate that the most important factor affecting capillary trapping is the different entry point pressures for the different geologic unit types (the so called secondary seal effect).

**Table 3**

| Case # | Permeability | Capillary pressure | Hysteresis | Relative Permeability tables | Number of tables |
|---|---|---|---|---|---|
| 1 | Heterogeneous | on | on | Different for sand and OFG | 12 |
| 2 | Homogeneous, Anisotropic | on | on | The same for all cells | 6 |
| 3 | Heterogeneous | on | on | The same for all cells | 6 |
| 4 | Heterogeneous | on | off | Different for sand and OFG | 6 |
| 5 | Heterogeneous | off | on | Different for sand and OFG | 6 |

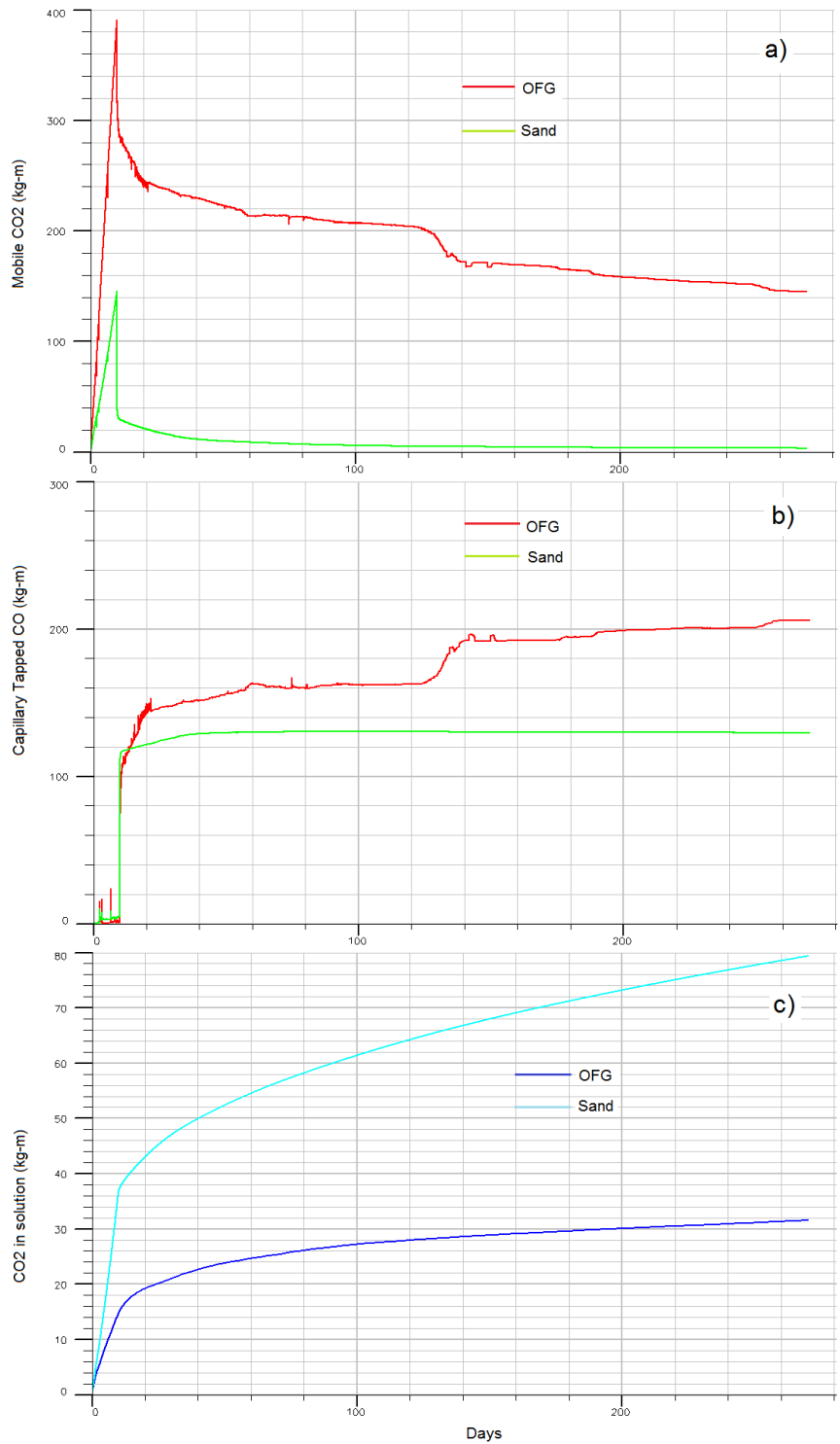

Figure 6. Total amount of the mobile (a), capillary trapped (b), and solute CO2 (c) in sand and OFG as function of time.

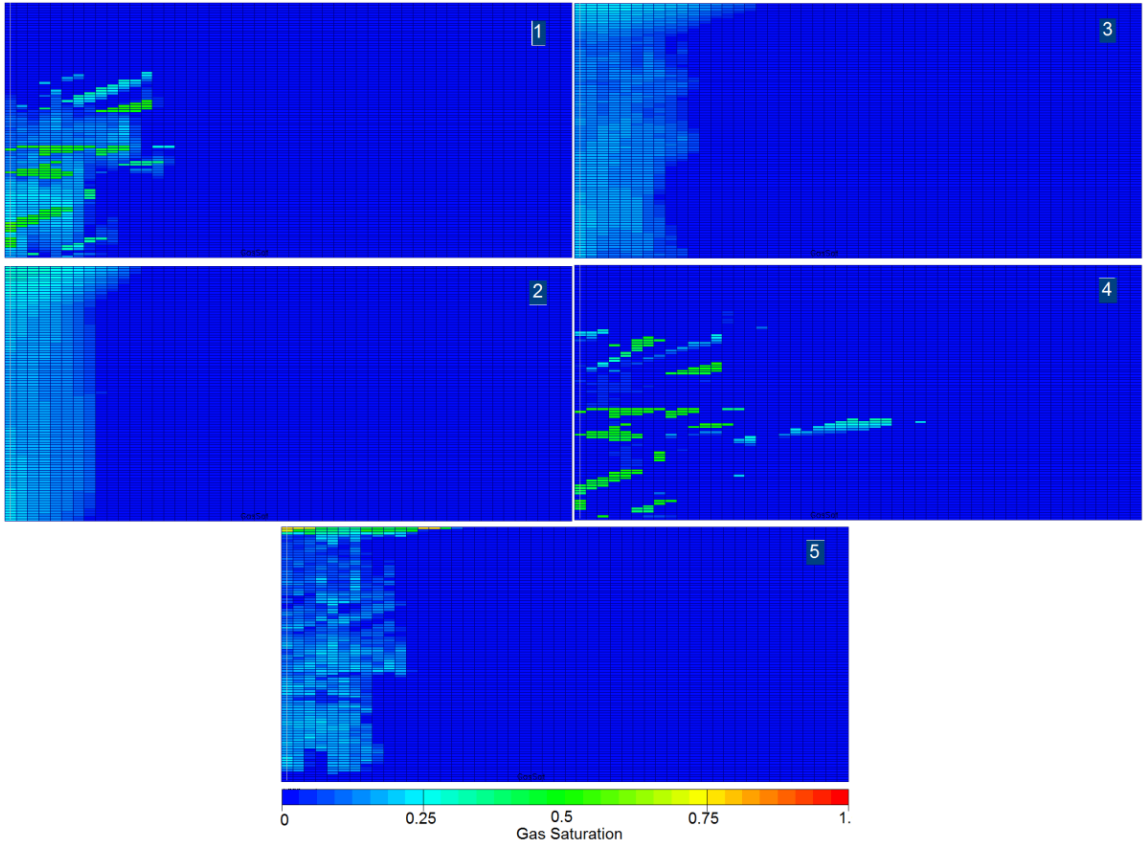

Figure 7. Vertical cross section of reservoirs showing CO2 saturation after 170 days from the beginning of CO2 injection for the five cases described in the Table 3. Well placed at a depth 2360 m.

## 6. CONCLUSIONS

Several features distinguish our approach from others:

1) The modeled heterogeneity structure and scales, hence permeability map, realistically reflects the typical fluvial type reservoirs; permeability ranges by 4 orders of magnitude.

2) The size of reservoir heterogeneities ranges from a few cm to dozens of meters.

3) The reservoir contains two different materials -- sand and open framework gravel -- with different properties, requiring that two sets of property tables be used for simulation.

4) Capillary pressure and hysteresis effects are utilized in the simulation. Overall, 12 properties tables were used including relative permeability and capillary pressure curves for drainage and imbibition for both brine and CO2.

The results demonstrate that small and medium scale inclusions of high-permeability cross strata fundamentally control trapping processes and hence the shape and dynamics of the CO2 plume. This occurs because the capillary entry pressures of the two materials are different. In a highly heterogeneous reservoir this trapping mechanism may considerably surpass all other capillary trapping mechanisms. Indeed, in the example considered here (Figure 2b) the total amount of injected CO2 is trapped and never reaches the top of the reservoir. The results strongly suggest that representing these small-scale features, and representing how they are organized within a hierarchy of larger-scale features, is critical to understanding trapping processes. It is also obvious that ignoring both the small-scale heterogeneity and the secondary-seal effect in simulations of CO2 sequestration may produce misleading results.

A number of studies have considered CO2 sequestration in fluvial reservoirs but have not represented sedimentary architecture at the scale of our study (e.g. Zhou et al., 2009; Delshad, et al, 2013). The results of our study indicate that the amount of trapping of CO2 and the geometry of the CO2 plume may be very different from the results of their current simulations if sedimentary architecture is represented at smaller scales.

The total amount of trapped (immobile) gas and its spatial distribution are different in heterogeneous and homogeneous cases with similar averaged characteristics. It is interesting that the mobile part of the gas is placed mostly inside of high-permeablity material, i.e. OFG. The ratio between amounts of gas in OFG to sand is about 8, although reservoir contains only 24% of OFG material. At the same time the amount of capillary trapped immobile gas is larger in sand.

Plume dynamics and the amount of trapped CO2 depend on the structure and content of the OFG cross strata, and how these are organized within larger units. The large OFG clusters and especially spanning clusters are responsible for the horizontal extent of CO2 plume, which may three times larger than in the case of a homogeneous reservoir. The size and shape of clusters are important with respect to the amount of gas trapped.

The simulations presented here were performed using a relatively small piece of the geologic model and with injection of small amount of CO2. We expect that simulations representing larger scales of the stratal architecture will be important to further understanding trapping in the reservoir. It is also important to study how the amount, rate, and schedule of CO2 injection affect trapping processes.

**ACKNOWLEDGMENTS**

We thank Schlumberger Limited for the donation of ECLIPSE Reservoir Simulation Software and the Ohio Supercomputer Center for technical support.